\documentclass{article}
\usepackage{arxiv}

\usepackage[utf8]{inputenc} 
\usepackage[T1]{fontenc}    
\usepackage{hyperref}       
\usepackage{url}            
\usepackage{booktabs}       
\usepackage{amsfonts}       
\usepackage{nicefrac}       
\usepackage{microtype}      
\usepackage{lipsum}
\usepackage{amssymb}

\usepackage{listings}
\usepackage{amsmath}
\usepackage{empheq}
\usepackage[most]{tcolorbox}

\usepackage[]{algorithm2e}

\usepackage{smartdiagram}

\usepackage{lineno}
\usepackage{amsmath}

\usepackage{amsthm}
 
\theoremstyle{definition}
\newtheorem{definition}{Definition}[section]

\title{Node Alertness \\ 
Detecting changes in rapidly evolving large graphs}

\author{
Mirco A. Mannucci \\
  HoloMathics, LLC\\
  \texttt{mirco@holomathics.com} \\
   \And
 Deborah Tylor  \\
  Tylor Data Services,  LLC
 \\
  \texttt{dtylor@tylordata.com} \\
}

\begin{document}

\maketitle

\begin{abstract}
In this article we describe a new approach for detecting changes in rapidly evolving large-scale graphs. The key notion involved is local alertness: nodes monitor change within their neighborhoods at each time step. Here we propose a financial local alertness application for cointegrated stock pairs.

\end{abstract}


\section{Evolving large scale Graphs}
\label{S:1}

\subsection{Background}
Graphs are becoming widespread in the Data Analysis community. Many real-life scenarios can be flexibly modeled via property graphs: nodes and edges may have attributes attached, thereby modeling concrete entities and their various types of connections (for instance humans and different relations between them:\textit{ Mirco is a colleague of Deborah's in the field of graph analytics, Deborah is a friend of Elena} ...).

Most realistic graphs evolve over time. They can be modeled by a family of graphs

\[
 \boxed{\{G_t\} = \{ (N_t, E_t) \}}
 \]

where $t$  is either a discrete or continuous time parameter. \footnote{notice in passing that this representation makes them effectively like generalized time series, except that instead of being numerical, their value at a particular time is a whole graph. The comparison might appear a bit far fetched, but the following may help seeing why it is not: suppose $G \xrightarrow{f} \mathbf{R}$ is a global function from a graph to the real numbers, for instance its diameter. Then, the composite $f(G_t)$ is an ordinary time series}

In case the graph $\{ G_t\}$ happens to be a property graph, it is understood that properties values (and sometimes even the properties list itself) may also change over time.

Due to graphs broad range of applications, it comes as no surprise that for many years  an entire industry has grown around graph-based solutions: graph databases (Neo4j, Titan, HypergraphDB, and others), graph processing platforms (Apache Giraph and Spark GraphX), graph visualization tools (GraphViz and CytoScape).

\subsection{Graph Mining: Relevance of Graphical Models}

\textbf{Graph Mining} is a relatively new discipline, born as a sub-chapter of Data Mining (about 10-15 years old, for an excellent early survey see the handbook on Graph Mining  \cite{DBLP:books/crc/p/GleichM16}). Graph Mining aims at finding patterns in graphs (a typical example is frequent sub-graphs, or, at the other side of the spectrum, very infrequent sub-graph), see the excellent survey by Faloutsos and others 
\cite{DBLP:series/synthesis/2012Chakrabarti} .

 Graph Mining has developed its own toolbox (e.g. methods to detect communities in social graphs). These algorithms often present a major challenge, as soon as one shifts focus from small and mid-size graphs to larger ones. The reason is that proper distributed versions of these techniques are not always available, and turning single thread graph processing algorithms into their distributed analogues requires judicious choices. \footnote{One reason is this: graphs cannot always be broken into a list of disconnected components of roughly the same size. Thus, more often than not, techniques following the parallel processing, map-reduce paradigm cannot be applied simply. }

\subsection{Alert Detection in big dynamic graphs }

In evolving graphs we are interested in detecting patterns of change, with the ultimate goal of prediction based on these changes. As we have already mentioned, an evolving graph is like a generalized time series, where the graph structure itself varies through time, beyond a simple number value variation. An example, which is relevant to our endeavor could be: assume you have a "health function" for the graph, which returns a value in $[0, 1]$ at each moment in time. Then the "graph health" is an ordinary time series, and we may ask for its moving average; or we may try to predict the graph's health at the next step on the basis of a windowed view of the graph series.

Although the area of evolving graph mining is fairly new, some methodologies have already emerged (see \cite{DBLP:conf/icdm/2008w} ). A typical strategy is to try to learn the patterns of changes over time, as a set of \textbf{graph production rules}. Unfortunately learning such patterns is quite computationally expensive, and, at least for now, not overly amenable to either real-time monitoring or prediction. 

Thus, more modestly, we may be trying to detect some relevant changes between two consecutive time ticks (see also \cite{DBLP:journals/ida/EberleH15}). These changes can be either local or global: in other words, we may be interested in anomalies at some local level, around a particular node or group of neighboring nodes,  or we are may be interested in the global health of the entire graph, with respect to a global health function. When structural anomalies are detected we wish to raise an alert.

\subsection{Node-centered graph computing }

One of the emerging distributed computing paradigms in large graphs processing is \textbf{vertex-centered computing}: nodes become  smart, and can send and receive messages with their neighbors; an event can thus propagate through the graph.  Communication is limited to each node's local neighborhood; nodes which belong to different connected components, cannot communicate with each other, either directly or indirectly. Nodes can then make informed decisions, after aggregating messages from their neighbors. 

Notice that this graph messaging paradigm can be accomplished in  different ways: on the one hand, it can be entirely asynchronous (each nodes receives/sends messages and updates its status in a completely independent fashion) or alternatively, follow a synchronous regime. 

Whereas there can be no doubt that the asynchronous model is more appealing, it also presents problems: it opens the gate to issues, such as deadlocking, inconsistent states, etc. Therefore the current prevalent paradigm, known as \textbf{Bulk Synchronous Parallel Model}, or\textbf{ BSP},  imposes some structure of synchronicity. 

Google has created an implementation of this paradigm known as \textbf{Pregel} (see \cite{DBLP:conf/sigmod/MalewiczABDHLC10}). Pregel is comprised of  \textbf{super-steps}: at each super-step all nodes process received messages, update their status, and send messages to be processed in the next iteration.How many iterations? At each super-step a node can elect to flag itself as done. The process terminates when either all nodes are done, or after a predetermined maximum number of super-steps.


\section{Node Alertness: detecting rapid change at the local level}
\label{S:2}

\subsection{Node Alertness Pipeline in a nutshell}
It now time to formulate our proposal: it boils down to performing a number of  super-steps after each graph update, during which nodes receive updates from their neighbors. Each node then decides, on the basis of custom built-in logic, whether the information received and integrated is worthy of an alert. The last phase of Node Alertness is an assessment of what to do with the local alertness signals. Here is the summary:

\begin{table}[ht]
\centering

\begin{tabular}{ |p{6cm}||p{6cm}|  }
 \hline

 \textbf{  STEP  } & \textbf{DESCRIPTION} \\
\hline
\hline
 \textbf{ PARTIAL/TOTAL UPDATE GRAPH }  & Update the graph via an interval of streaming \\ 
 
  \hline
\textbf{MESSAGE  NEIGHBORS} & Nodes send messages to their neighbors \\
 \hline
\textbf{NODE EVALUATES ALERT STATUS} & Each node processes messages received  \\
 \hline
\textbf{ALERTS INTEGRATION AND REPORT} & Integrate the local alerts \\
 \hline
\end{tabular}
\caption{Steps in Node Alertness}
\end{table}

\begin{enumerate}
\item In the first step, we update the entire graph (notice that we could update both nodes and edges, which entails that even the topology of the graph may change). The update represents all changes in a single time click. 
\item We execute one or more   \textbf{ super-steps} in Pregel's jargon. Each node send its updates to all its neighbors (again, its neighbor may have changed from the last super-step)
\item At each super-step, every node process messages and decides whether something meaningful has happened (here of course the detail depends on the context: sometimes the node simply detects that some function whose value depends on all statuses of neighbor has gone beyond a certain threshold). If node decides that something is anomalous, it can either raise a local alert and flag itself as done or remain open for more information. 

\item After the sequence of super-steps, each node in the graph has either raised an alert or not. We can simply collect all nodes alerts and decide, on the basis of some global health function, whether it is worthwhile to raise a global alert. 
\end{enumerate}

In this schema we have deliberately omitted two critical pieces:

\begin{enumerate}
\item The message/update function that provides the "brain" of each node. In other words, the embedded logic that turns each node into a mini-processing machine

\item The global reduce/integrate function which takes the entire graph, reads nodes new statuses, and return a global graph alert. 
\end{enumerate}

The motive behind such a glaring omission is this: neither of the two functions can be decided in the absence of outside domain knowledge. Furthermore, sometimes we may be happy with assessing local alerts, without a global picture. In this case, what we may want to do is implement a local response to the local alert (see the last section). 

\subsection{Node Alertness in Apache Spark}

Apache Spark is the new star in the Big Data Computing arena.  After a decade long kingship of Hadoop, many organizations are migrating steadily to Spark. Why? Spark allows analysts to carry out computations in distributed memory. Unlike the batch processing approach of Hadoop, Spark makes for interactive distributed computations, something very useful in the context of data mining (instead of doing everything at once, the analyst can interact and guide the computing machine toward a new goal through a pipeline of steps).

Spark is equipped with four capabilities that enable Node Alertness:

\begin{enumerate}
\item \textbf{STRUCTURED STREAMING} a tool to handle streaming data from many sources (see \cite{DBLP:conf/ipps/ChintapalliDEFG16}
)
\item \textbf{MLIB}, a library for Machine  Learning
\item \textbf{GRAPH FRAMES} an add-on which uses Spark Dataframes to build and manipulate graphs. It includes built-in functions to implement Pregel's style  graph computing
\item\textbf{Databrick's DELTA} a unified analytic engine and associated table format built on top of Spark. It provides ACID transactions, indexes for building data pipelines to support big data use cases, from batch and streaming ingests, fast interactive queries to machine learning.

\end{enumerate}

Now, let us see how we can leverage Spark's resources to implement a \textbf{Node Alertness Pipeline} which follows the schema of the previous section.

To begin with, we use Spark Structured Streaming to ingest from a target source new data, in time clicks chunks. There are already a number of ingestors available out of the box, but if we plan to hook our machinery to some other source,  we can implement our own stream ingestor and register it with Streaming.

Now, we initialize the graph once, as a graphframe.  At each successive time click we simply update the graph to incorporate changes.The update triggers a single Pregel job, for messaging and local processing. Finally, if we need some global alertness, we can leverage a reduce function which is part of the Spark arsenal. 

We could have several super-steps for a single update in some scenarios. For instance, suppose that each node checks its immediate neighbors in the first super-step. Finding nothing anomalous, it declares itself as done. If, on the other hand, an anomaly in its immediate neighborhood is detected, it may be enough to raise an alert, or the node may opt to know what \textit{each} of its neighbors have found in their neighborhood. In this case a second super-step might be required. In general, the more super-steps are allowed around each update, the more problems are introduced. As a rule of thumb, for real-time or next to real-time applications, it is wise to keep the number of acceptable super-steps to a bare minimum, ideally one. This is precisely what we shall do in the experiment of the next section.

\section{Node Alertness in Financial Forecasting}
\subsection{Cointegration graphs}
Although the methodology which we have described applies to virtually all large dynamic graphs, we would like to showcase a financial application which, aside its inherent value, can also be generalized to other  financial monitoring scenarios. In his 2017 ForecastNYC talk, Michael Kane introduced a specific graph, that could be called \textbf{Co-Integration Financial Graph}: nodes are financial symbols, say stocks names listed in the NASDAQ, whereas the  edges represent co-integrated pairs of  symbols over a certain time window (see \hyperlink{https://presagia.slides.com/michaelkane/a-cointegration-approach-to-identifying-systemic-risk-in-markets}{Michael Kane's slides here}). Just for convenience's sake, we recall here the definition of co-integration:

\theoremstyle{definition}
\begin{definition}{\textbf{COINTEGRATION OF TIME SERIES }}

Two non-stationary time series $x_t$ and $y_t$ are said to be co-integrated if 
$y_t - \beta * x_t  = u_t$ where $u_t$ is stationary. 
\footnote{co-integration of time series aims at capturing their correlation but not point-wise: rather, it is like having a random walk with your dog. The dog may stray a bit, but not too far, because of the leash. see \cite{murray1994drunk} for a very pictorial way to get the right intuition about co-integration.

}

\end{definition}

To avoid having a completely connected graph, we can set a convenient threshold: if cointegration p value is below an assigned threshold, we draw no edge between two symbols.

\theoremstyle{definition}
\begin{definition}{\textbf{COINTEGRATION GRAPH}}

Given a family of times series $FTS= \{ T_t , S_t, \ldots\}$ and a threshold $\epsilon$, the   associated co-integration graph $ G(FTS, \epsilon)$ is a directed  graph  whose nodes set is $FTS$ and edges are established for every pair $(S_t, T_t)$ such that the $p$ value of the cointegration test is below  $\epsilon$

\end{definition}

\begin{figure}[ht]
\centering\includegraphics[width=0.7\textwidth]{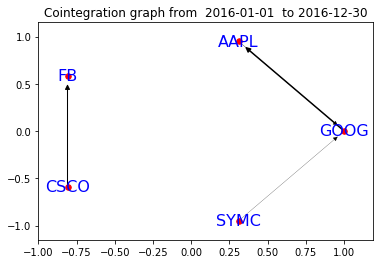}
\caption{Co-Integration Graph of  symbols GOOG, AAPL, FB, INTC, CSCO, PYPL, SYMC}
\end{figure}

In Figure $1$ we show a sub-graph of the full SP 500 cointegration graph at some point in time for only a few symbols. The edges thickness is inversely proportional to the standard deviation of the residual of the co-integration  test. In other words,the smaller the standard deviation, the stronger the tie is between the nodes.

As the window shifts , the graph changes. 

\begin{figure}[ht]
\centering\includegraphics[width=0.7\textwidth]{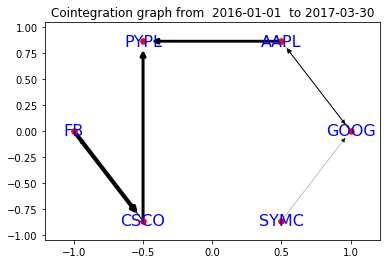}
\caption{Cointegration graphs of the same symbols, but with shifted time window}
\end{figure}

Figure $2$ shows the evolution of the sub-graph above. Notice that some symbols have strengthened their co-integration degree.

In this scenario  we have swept under the rug an important point: if we wish to recalculate the entire graph at each time tick we should recompute the cointegration of all node pairs, an expensive thing to do, even with the power of cloud computing. Secondly, it would hardly make  sense, as the co-integration test requires a larger time horizon. 

Rather than continually recalculate the cointegration graph, we propose an alternative: as each node updates its ticker value through time, it then checks whether each of the neighboring nodes, found in the cointegration graph, are within a reasonable distance given their new values and assuming their pairwise cointegration has not changed (details in the next sub-section).

Using the Node Alertness Pipeline advocated here, one can develop an entire gamut of financial monitoring systems. For instance, one can add a global function that stipulates the following: if too many nodes within a certain sector raise an alert, declare a global alert. Or perhaps we may decide that it is fine for many nodes in risky sectors to cluster up, but if alertness spreads across sectors it is a signal of real trouble. It is easy to see that the pipeline is  flexible enough to accommodate numerous scenarios,  all based on the same graph. 

\subsection{An experiment: when stock pairs break the leash.}
We describe now a simple experiment in Python which highlights the methodology we have introduced thus far.

Rather than monitoring whether cointegration has become tighter for some cluster of nodes, we try to detect whether some nodes \textit{have broken their leash}. In other words, assuming the hypothesis that the underlying cointegration model has not changed, we will identify pairs of nodes whose new values are too far apart with respect to their expected range. The ones which are now too far apart are those who have broken the leash, i.e. the cointegration model has changed. \footnote{Without entering an area which is beyond the scope of this article, we notice \textit{en passant} that this information can be extremely useful for operators dealing with pair trading. If we know that two stocks are no longer cointegrated or the underlying model has changed, we can leverage this information to redefine or adjust strategy (more on this in a follow up article). } 

We have ingested a data set of symbols and their prices over a time window. The following details the construction of the corresponding graph:

\begin{algorithm}[H]
 \KwData{SP 500 windowed data   }
 \KwResult{SP 500 co-integration graph }
 initialization\;
 \While{not at end of data set}{
 1 calculate linear regression for each pair of stocks\;
 2 save the model for the pair ( intercept $\beta_0$ , slope $\beta_1$, as well as  means $\mu$ and standard deviation $std$ of residuals)\;
 3 determine whether the residuals series is stationary via the AdFuller test\;
 4 save p value for the pair\;
  }
  
 \While{not at end of dataframe}{
 1 create GraphFrame edge if p value is below a critical threshold (default used 0.05)\;
 2 save model parameters and standard deviation as attributes of the edge\;
 
  }
  \caption{Constructing SP 500  cointegration graph over an assigned time window}
\end{algorithm}

After building the co-integration graph, at each new time tick we update the attribute of each node \footnote{As a graphframe is comprised of two dataframes, one for the nodes and one for the edges, we simply construct a new graphframe whose nodes dataframe has been created by the update. }. Pregel sends the new values of all neighboring nodes to a given node. Its logic is simple: it checks whether the  new residual using the model is within three standard deviations of its means. If it isn't, a local alert is thrown:

\begin{algorithm}
 \KwData{Co-integration graph  }
 \KwResult{Graph local status }
 initialization\;
 \While{new data chunk}{
 1 update node values with last stock closing price
 \;
 2 begin pregel job: nodes send their current prices to their neighbors\;
 3 each node checks the values of all its neighbors. If actual  price is beyond three standard deviations of predicted value it raises an alert\;
  }
 \caption{Node Alertness Loop}
\end{algorithm}

 Using a simple python script for demonstration (\cite{GitHubNodeAlertness2019}), we have calculated the entire SP 500 co-integration graph over period of the January 2015 to December 2015. 
  In Figure 3, we see a small portion of the SP500 cointegration graph whose nodes are all highly co-integrated pairwise. It is a cluster of 64 nodes, comprising the following symbols: 
  
'NKTR', 'HSIC', 'DD', 'CNC', 'CELG', 'WU', 'AKAM', 'ROST', 'PM', 'CMS', 'WM', 'EQR', 'CCI', 'MSFT', 'DRI', 'ALXN', 'DLR', 'CB', 'UPS', 'SIVB', 'XEC', 'LOW', 'FTNT', 'AIV', 'GE', 'JNJ', 'LRCX', 'ADSK', 'MTD', 'TRV', 'KLAC', 'AEE', 'NI', 'SYF', 'ROP', 'T', 'ED', 'BAX', 'KO', 'ABBV', 'TXN', 'XEL', 'CMG', 'AMT', 'INTU', 'XYL', 'LUV', 'INTC', 'GILD', 'DG', 'PAYX', 'REG', 'CBRE', 'MET', 'WCG', 'GS', 'HCA', 'ADI', 'CPRT', 'IPG', 'CHD', 'PRGO', 'BBT', 'FRT''
  
  We calculated how many pairs had anomalous behavior on January 20th 2016. Out of $420$ only $310$ survived and were compatible with the cointegration model. It should be pointed out that the day chosen was a turbulent one. 
 
\begin{figure}[ht]
\centering\includegraphics[width=0.7\textwidth]{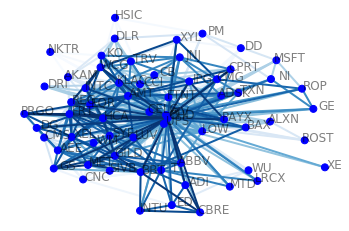}
\caption{Highly cohesive subgraph of SP cointegration graph }
\end{figure}
Figure 4 is the same subgraph where edges which are broken on January 20 2016 have been removed:

\begin{figure}[ht]
\centering\includegraphics[width=0.7\textwidth]{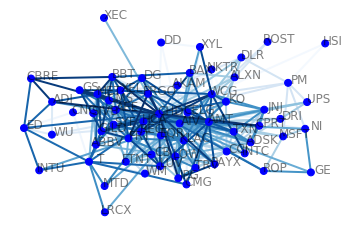}
\caption{Same subgraph pruned of broken edges. }
\end{figure}

Raising a local alert could  also prompt  a  \textbf{selective update of the graph} as a side effect. In other words, instead of recomputing the co-integration test for the entire graph, we can now trigger the recalculation only for pair of nodes who have anomalous behavior. The advantage of this approach is that, assuming a relatively sporadic and infrequent set of local alerts, such a surgical re-computation of the co-integration graph is still within the compass of near real time monitoring systems.

\section{Future work in Node Alertness}
\label{S:3}
In this last section we list a few directions of research which expand the scope of Node Alertness.

\subsection{Toward smart resilient large scale graphs}
A natural step beyond raising alerts would be some kind of local  graph intelligence ensuring global resilience of the network. What this means is that each  component of the graph takes care of itself, i.e. reacts to changes in some clever way to ensure stability. This line of reasoning seems to be quite fruitful in the financial word: just like the cointegration graph which we have mentioned can be coupled with several monitoring pipelines, other financial graphs could be equipped with similar alerts and responses mechanism to establish resilience (we are now thinking here of using this monitoring for reassessing joint symbols trading strategies).

\subsection{ Local Alertness with memory }

The approach detailed in the previous section is a bit simple-minded: nodes have a just a boolean alert status.  If we need to make them a bit smarter, for instance being able to raise different levels of alert on the basis of previous alert statuses, we need to provide them with some kind of memory.  How? As we are operating in Spark's world, there are essentially two main options:
\begin{enumerate}
\item each node comes equipped with a local memory,a map whose key is a time stamp and its value is an alert state. At each super-step, the node consults the map up to now, makes its decision, and updates it inserting  the last status. 
\item a shared memory, accessible to all nodes. This can be accomplished in Spark by the so-called broadcasting: some data are shared among all instances. Of course, this trick works well as long as the share-able data is not too big, else overall  performance of the cluster will be eroded.

\end{enumerate}

\subsection{From  Local Alertness to prediction.}

As we said in the first section, we have restricted ourselves to alertness, refraining for the time being from any attempt to leverage this information for prediction. But, in conjunction with the methods sketched in the previous subsection, we can indeed collect data on how local alerts spread over several time ticks. In turn, this data may be useful to both categorize the current status of the graph as a whole, and to predict further changes. In a paper in preparation (\cite{TylorMannucci2019}  ) we shall show how the growth of local alerts and their diffusion pattern can be instrumental in assessing the health of a co-integration graph, for instance of the market as a whole, or sectors thereof.


 \appendix


\bibliographystyle{unsrt}  \bibliography{references}  

\begin{thebibliography}{1}

\bibitem{DBLP:books/crc/p/GleichM16}
David~F. Gleich and Michael~W. Mahoney.
\newblock Mining large graphs.
\newblock In Peter B{\"{u}}hlmann, Petros Drineas, Michael Kane, and Mark~J.
  van~der Laan, editors, {\em Handbook of Big Data.}, pages 191--220. Chapman
  and Hall/CRC, 2016.

\bibitem{DBLP:series/synthesis/2012Chakrabarti}
Deepayan Chakrabarti and Christos Faloutsos.
\newblock {\em Graph Mining: Laws, Tools, and Case Studies}.
\newblock Synthesis Lectures on Data Mining and Knowledge Discovery. Morgan
  {\&} Claypool Publishers, 2012.

\bibitem{DBLP:conf/icdm/2008w}
{\em Workshops Proceedings of the 8th {IEEE} International Conference on Data
  Mining {(ICDM} 2008), December 15-19, 2008, Pisa, Italy}. {IEEE} Computer
  Society, 2008.

\bibitem{DBLP:journals/ida/EberleH15}
William Eberle and Lawrence~B. Holder.
\newblock Scalable anomaly detection in graphs.
\newblock {\em Intell. Data Anal.}, 19(1):57--74, 2015.

\bibitem{DBLP:conf/sigmod/MalewiczABDHLC10}
Grzegorz Malewicz, Matthew~H. Austern, Aart J.~C. Bik, James~C. Dehnert, Ilan
  Horn, Naty Leiser, and Grzegorz Czajkowski.
\newblock Pregel: a system for large-scale graph processing.
\newblock In Ahmed~K. Elmagarmid and Divyakant Agrawal, editors, {\em
  Proceedings of the {ACM} {SIGMOD} International Conference on Management of
  Data, {SIGMOD} 2010, Indianapolis, Indiana, USA, June 6-10, 2010}, pages
  135--146. {ACM}, 2010.

\bibitem{DBLP:conf/ipps/ChintapalliDEFG16}
Sanket Chintapalli, Derek Dagit, Bobby Evans, Reza Farivar, Thomas Graves, Mark
  Holderbaugh, Zhuo Liu, Kyle Nusbaum, Kishorkumar Patil, Boyang Peng, and Paul
  Poulosky.
\newblock Benchmarking streaming computation engines: Storm, flink and spark
  streaming.
\newblock In {\em 2016 {IEEE} International Parallel and Distributed Processing
  Symposium Workshops, {IPDPS} Workshops 2016, Chicago, IL, USA, May 23-27,
  2016}, pages 1789--1792. {IEEE} Computer Society, 2016.

\bibitem{murray1994drunk}
Michael~P Murray.
\newblock A drunk and her dog: an illustration of cointegration and error
  correction.
\newblock {\em The American Statistician}, 48(1):37--39, 1994.

\bibitem{GitHubNodeAlertness2019}
Deborah~A. Tylor and Mirco~A. Mannucci.
\newblock Node alertness.
\newblock \url{https://github.com/dtylor/nodealertness.git}, 2019.

\bibitem{TylorMannucci2019}
Deborah~A. Tylor and Mirco~A. Mannucci.
\newblock Pattern mining in cointegration graphs.
\newblock Working paper, 2019.

\end{thebibliography}



\end{document}